\documentclass{article}  
\usepackage{url}
\usepackage{graphicx}
\usepackage[show]{ed}
\usepackage[hyperref,backend=bibtex,style=alphabetic]{biblatex}
\usepackage{bibtweaks}
\usepackage{hyperref}
\title{The Theorem Prover Museum -- 
Conserving the System Heritage of Automated Reasoning}
\author{Michael Kohlhase\\
Computer Science, FAU Erlangen-N\"urnberg\\\url{http://kwarc.info/kohlhase}}
\begin{document}
\maketitle
\begin{abstract}
  We present the Theorem Prover Museum, and initiative to conserve -- and make publicly
  available -- the sources and source-related artefacts of automated reasoning systems.
  Theorem provers have been at the forefront of Artificial Intelligence, stretching the
  limits of computation, and incubating many innovations we take for granted
  today. Without the systems themselves as preserved cultural artefacts, future historians
  will have difficulties to study the history of science and engineering in our
  discipline.  
\end{abstract}
\section{Introduction}\label{sec:intro}

Theorem provers are software systems that can find or check proofs for conjectures given
in some logic. Research in theorem proving systems started with Newell and Simon's ``logic
theorist'' 1955~\cite{NewSim:ltmcips56} -- one of the earliest systems in the
then-emerging field of Artificial Intelligence -- and has led to a succession of systems
since. Today, more than 60 years later, the CADE ATP system competition~\cite{CASC}
attracts 15-20 systems annually. Automated reasoning systems have applications ranging
from the verification of mathematical results, via program synthesis/verification, the
Semantic Web, all the way to the discovery of unfair trading rules in darkpools of
investment banks.

Theorem provers are complex software systems that have pushed the envelope of artificial
intelligence and programming, and as such they constitute important cultural artefacts
that carry within them the beginnings of many aspects of computing we take for granted
today. To name just one example: the programming language ML: (Proof) Meta-Langauge which
heavily influenced modern typed functional programs was introduced as a meta-language of
the LCF theorem prover by Robin Milner. Its type system was motivated by the idea that
proofs could be programmed, if the type of proofs can only contain logically valid
proofs.

With the ongoing wave of retirements of the original principal investigators there is good
chance that these systems are lost, when their group servers are shut down. The following
incident is unfortunately quite typical. When -- ten days after Herbert Simon's passing in
February 2001 -- the author tried to find a copy of the source code of the Logic Theorist
in Simon's scientific estate at CMU, all tapes and printouts had already been discarded --
only the written materials and notes were being catalogued in the CMU
library. Fortunately, report P-868 of the Rand Corporation~\cite{NewSim:ltmcips56}, where
the program was conceived contains the full printout of the code. Otherwise we would only
be able to read about this seminal program, but not be able to study the artefact
itself. 

In other cases, we may not have been so lucky; see~\cite{tpmuseum:tpbl:on} for a list of
theorem provers believed lost. This is a great loss to the culture of our discipline,
which is in danger of becoming marginalized by the hype waves rolling through AI and
computing. Without the systems as preserved cultural artefacts, future historians will
have difficulties to study the history of science and engineering.

\section{A Museum of Theorem Prover Source Code and Artefacts}

This article reports on an initiative started by the author in spring 2016 to help
conserve the source code of theorem provers: the ``theorem prover museum'', a collection
of GitHub repositories with source code of systems, together with a web site that presents
them and organizes the process of acquiring more.

The term ``museum'' in the title may sound a bit ambitious, since the exhibition and
didactic interpretation of the theorem provers is beyond the scope of the initiative (and
perhaps abilities of the founder). But the foremost function of any museum is the
conservation of artefacts, which is what the ``theorem prover museum'' project intends to
do. Once the source code is preserved, historians of science and engineering can start to
do research on it and create multiple user interfaces to present it to the public.

Note that it is not the purpose of the museum to keep the theorem proving systems running
(in many cases the compilers and dependencies have moved on, making this very
difficult). But only to archive the source code for academic study.  This is a
well-considered design decision, taken to lower the barrier of archiving systems
here. Again, once the source code is preserved -- i.e. made public by the original authors
-- other enthusiasts can possibly revive it. Indeed this has already happened, triggered
by the act of exposing the source in the museum.

\section{Realizing the Museum}

\begin{figure}[ht]\centering
  \includegraphics[width=\textwidth]{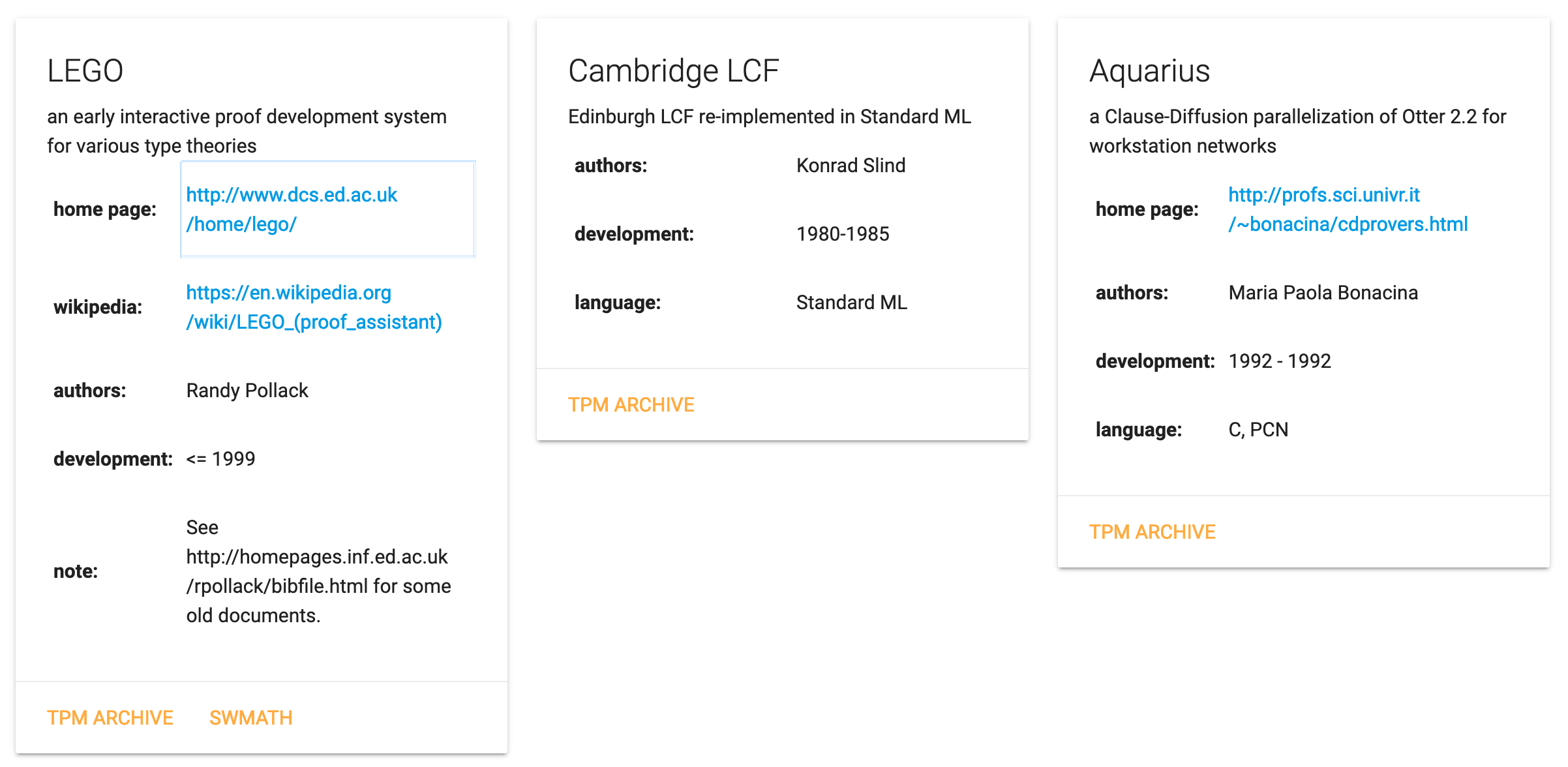}
  \caption{Three Theorem Prover Cards in the Museum}\label{fig:cards}
\end{figure}

The actual ``theorem prover museum'' consists of a simple web site at
\url{https://theoremprover-museum.github.io/} that features a couple of with cards with
short profiles for theorem provers (see Figure~\ref{fig:cards}) depending on their museum
status. The front page of the museum is the index of museum systems, i.e. systems that are
no longer actively maintained but for which a code repository exists. The repositories are
collected in the GitHub organisation \texttt{theoremprover-museum}
\url{https://github.com/theoremprover-museum}. An increasing number of systems already
have repositories (git or other), here we are working towards automatically maintaining a
local mirror repository in the museum -- just to keep the systems safe. 

Additionally the museum contains various administrative pages that collect systems, e.g. a
list of ``most wanted systems'', a list of ``theorem provers believed
lost''~\cite{tpmuseum:tpbl:on}, and a list of ``active systems''. Once in a while, a
request for the source code of a system that has fallen below the radar of the community
is met with an exasperated reply like ``but Ontic lives!!!'' (David McAllister in
2016).

All of these pages are statically generated from a central data
file \texttt{provers.yml} \cite{tpmuseum:data:on} which keeps nested key/value data in
YAML. This file can be extended by a simple pull request and has proven a low-maintenance
solution. 
    
Since the initiative was started, the museum has gained the source code of 38 systems,
which form a cross-section of the discipline. The systems span a period of 50 years, and
the code ranges from machine language to high-level languages like OCaml. Even though the
museum has some of the iconic systems of the field -- along with some of the more obscure
ones, it does not -- unfortunately -- constitute a fully representative sample yet. More
contributions and hunting down system sources is still needed for that.

The concept of the theorem prover museum is compatible with the Software Heritage
Initiative~\cite{SoftwareHeritage:on}, and particular GitHub-based implementation
contributes to it automatically, since the SHI indexes GitHub repositories and the museum
adds content that was unreachable to the SHI before.

The \textsf{swMath} information system for mathematical software~\cite{swMath:on} lists
the museum as one of its special categories~\cite{swMath:tpmuseum:on}. This links systems
to their traces in the mathematical literature -- unfortunately, much of the theorem
proving literature is in Computer Science conferences, which are only partially tracked in
the underlying \textsf{zbMATH} abstracting service~\cite{zbMATH:on}, but CS does not have
a comparable system. Even so, the \textsf{swMath} pages provide valuable additional
information for the museum systems. 

\section{Related Initiatives and Resources}
We list other public resources that may give further information
\begin{itemize}
\item there is a small literature on the history of automated reasoning, it
  includes~\cite{Bibel:ehpad07} on the early history up to 1970 and~\cite{RobVor:hoar01}
  for the next 30 years. 
\item the Encyclopedia of Proof Systems~\cite{Wolzenlogel-Paleo:teps17} collects proof
  systems that are mechanized by the theorem provers.
\item the Wikipedia pages on automated theorem provers and proof assistants keep list of systems
\item the program verification and synthesis community keeps a systems
  list~\cite{vsstp:on} that also contains a section on theorem provers.
\end{itemize}

\section{Conclusion \& Call for Contributions}\label{sec:concl}

We have presented an initiative for conserving the sources of historic theorem proversystems, i.e. systems that are no longer actively developed and in danger of loss. The
theorem prover museum is now fully functional as a system and has attracted various
entries. Even though it has been well received, it needs contributions from the community:
curators who chase down sources, talk to retired researchers who might know about the
whereabouts of source code, and even go to the basement and lug up dusty magnetic
tapes. In short the Indiana Jones types of Automated Reasoning -- without the ``stealing
from indigenous cultures'' part.

But most importantly, we need the individual researchers who, when they realize that they
have moved on from a project to routinely submit to the theorem prover museum just as we
submit a paper to a journal. The theorem prover museum gives them a place to do this and
thus to contribute to the immaterial legacy of our research field. 

\paragraph{acknowledgements}
I am grateful to many colleagues from the automated reasoning community, amongst all
contributors I would like to single out William Farmer, who submitted the first prover:
IMPS to the museum, Tom Wiesing who helped me with the web page, J\"org Siekmann and
Wolfgang Bibel, who were supportive to the idea from the inception, Mike Gordon and Konrad
Slind who chased down early versions of the HOL provers, and finally Rany Pollack, who
after enduring more than a dozen reminder finally dug up the LEGO source code and
contributed it.

\printbibliography
\end{document}